\documentclass[fleqn,12pt,twoside]{article}
\usepackage[headings]{espcrc1}

\readRCS
$Id: sorensen\_qm2005.tex,v 1.2 2004/02/24 11:22:11 spepping Exp $
\usepackage{graphicx}
\usepackage[figuresright]{rotating}

\newcommand{\AmS}{{\protect\the\textfont2
  A\kern-.1667em\lower.5ex\hbox{M}\kern-.125emS}}

\title{Elliptic flow fluctuations in Au+Au collisions at
  $\sqrt{s_{_{NN}}}=200$~GeV }

\author{P. Sorensen\address[MCSD]{Brookhaven National Laboratory, 
    Upton, New York 11973-5000, USA} for the STAR Collaboration\thanks{For the full list of STAR authors and acknowledgements, see appendix `Collaborations' of this volume} }%
       
\runtitle{$v_2$ fluctuations}
\runauthor{P. Sorensen}

\begin{document}
\maketitle

\begin{abstract}
  Please note that after these results were reported at Quark Matter
  2006 and posted on the preprint server it was found that what is
  reported here as \textit{elliptic flow fluctuations}, should rather
  be taken as an upper limit on the fluctuations. Further analysis has
  shown that fitting the multiplicity dependence of the q-distribution
  does not enable one to disentangle non-flow and fluctuations. The
  data from the q-vector distrubution does not, therefore, exclude the
  case of zero fluctuations. The remainder of these proceedings we
  leave as they were originally reported.\\ 
  We report first
  measurements of elliptic flow ($v_2$) fluctuations for the STAR
  collaboration at middle rapidity in $\sqrt{s_{_{NN}}}=200$~GeV
  Au+Au collisions. We analyze the multiplicity dependence of the
  flow vector length distribution to disentangle non-flow
  correlations from $v_2$ fluctuations. We find that the width of
  the $v_2$ distribution is approximately 36\% of the mean $v_2$
  and, within errors, independent of collision centrality. This
  value coincides with eccentricity fluctuations, apparently leaving
  little room for other sources of fluctuations.
\end{abstract}


\textit{Introduction:} A primary goal of the RHIC program is to
collide heavy-ions to create a quark-gluon plasma (QGP) and study its
properties. Measurements of the azimuthal anisotropy of particle
production with respect to the reaction plane (\textit{i.e.}
$v_2$)~\cite{v2papers} seem to indicate that the matter created in
collisions at RHIC behaves as a perfect fluid with a viscosity near a
conjectured lower bound~\cite{Kovtun:2004de}. This conclusion is based
primarily on comparisons to hydrodynamic model
predictions~\cite{hydro,v2papers}. Uncertainty about the conditions at
the beginning of the hydrodynamic expansion, however, make the
conclusion that the matter created at RHIC has a viscosity near the
conjectured lower bound ambiguous~\cite{cgcecc}. Since $v_2$ reflects
the initial spatial eccentricity of the overlap region when two nuclei
collide, fluctuations of $v_2$ should reflect fluctuations in the
initial eccentricity. As such, $v_2$ fluctuations may provide
sensitivity to the initial conditions and help determine the viscosity
and other properties of the matter created in heavy-ion collisions.


\textit{Analysis method:} The $n^{th}$ harmonic reduced flow vector is
defined as $q_{n,x} = \frac{1}{\sqrt{M}} \Sigma_i\cos(n\phi_i)$ and
$q_{n,y} = \frac{1}{\sqrt{M}} \Sigma_i\sin(n\phi_i)$, where $M$ is the
number of tracks and $\phi_i$ is the azimuth angle of a track with
respect to the reaction plane~\cite{qdist}. When enough tracks are
used in the calculation of $q_n$, the central-limit-theorem (CLT)
ensures that its distribution will be a two-dimensional Gaussian
shifted by $\sqrt{M}v_n$ in the x direction with widths:
\begin{equation}
 \sigma_x^2 = \frac{1}{2}\left(1 + v_{2n} - 2v_n^2 + g_n\right),~~~~~~~~  \sigma_y^2 = \frac{1}{2}\left(1 - v_{2n} + g_n\right).
\end{equation}
$g_n$ represents the broadening of the distribution that arises from
non-flow correlations~\cite{qdist}. The exact direction of the
reaction plane is not known, so we calculate the magnitude of the flow
vector $|q_n|$. If $v_2$ does not fluctuate, the distribution of $|q|$
should be given by:
\begin{equation}
\frac{dN}{q_ndq_n} = \frac{1}{\sqrt{\pi}\sigma_x\sigma_y}e^{-\frac{1}{2}\left(\frac{q_n^2+Mv_n^2}{\sigma_x^2}\right)}\sum_{k=0,2,4,...}^{\infty}\left(1-\frac{\sigma_x^2}{\sigma_y^2}\right)^k\left(\frac{q_n}{v_n\sqrt{M}}\right)^k\frac{1}{k!}\Gamma\left(\frac{2k+1}{2}\right)I_k\left(\frac{q_nv_n\sqrt{M}}{\sigma_x^2}\right) 
\label{dq}
\end{equation}
where $\Gamma$ is the gamma function and $I_k$ are modified Bessel's
functions. Here we consider the $n=2$ case. If $v_2$ fluctuates from
event to event, the final $|q_2|$ distribution will have the form:
$\frac{d\tilde{N}}{q_2dq_2} = \int dv_2 f(v_2) \frac{dN}{q_2dq_2}$,
where $f(v_2)$ is the probability for an event to have a given
$v_2$. Zero $v_2$ fluctuations corresponds to $f=\delta\left(v_2 -
\langle v_2 \rangle\right)$ where $\delta$ is the Dirac delta
function. In our analysis we assume a Gaussian shape for the $v_2$
distribution:
$f=\frac{1}{\sqrt{2\pi\sigma_{v2}^2}}\exp\left(-\frac{(v_2 -
  \langle v_2 \rangle )^2}{2\sigma_{v2}^2}\right)$, and extract
the r.m.s. ($\sigma_{v2}$) of the distribution from fits to
data. Different assumed shapes change the r.m.s. by at most 15\%
(relative).

\vspace{-0.4cm}
\begin{figure}[htb]
  \hspace{1cm}
  \resizebox{0.425\textwidth}{!}{\includegraphics{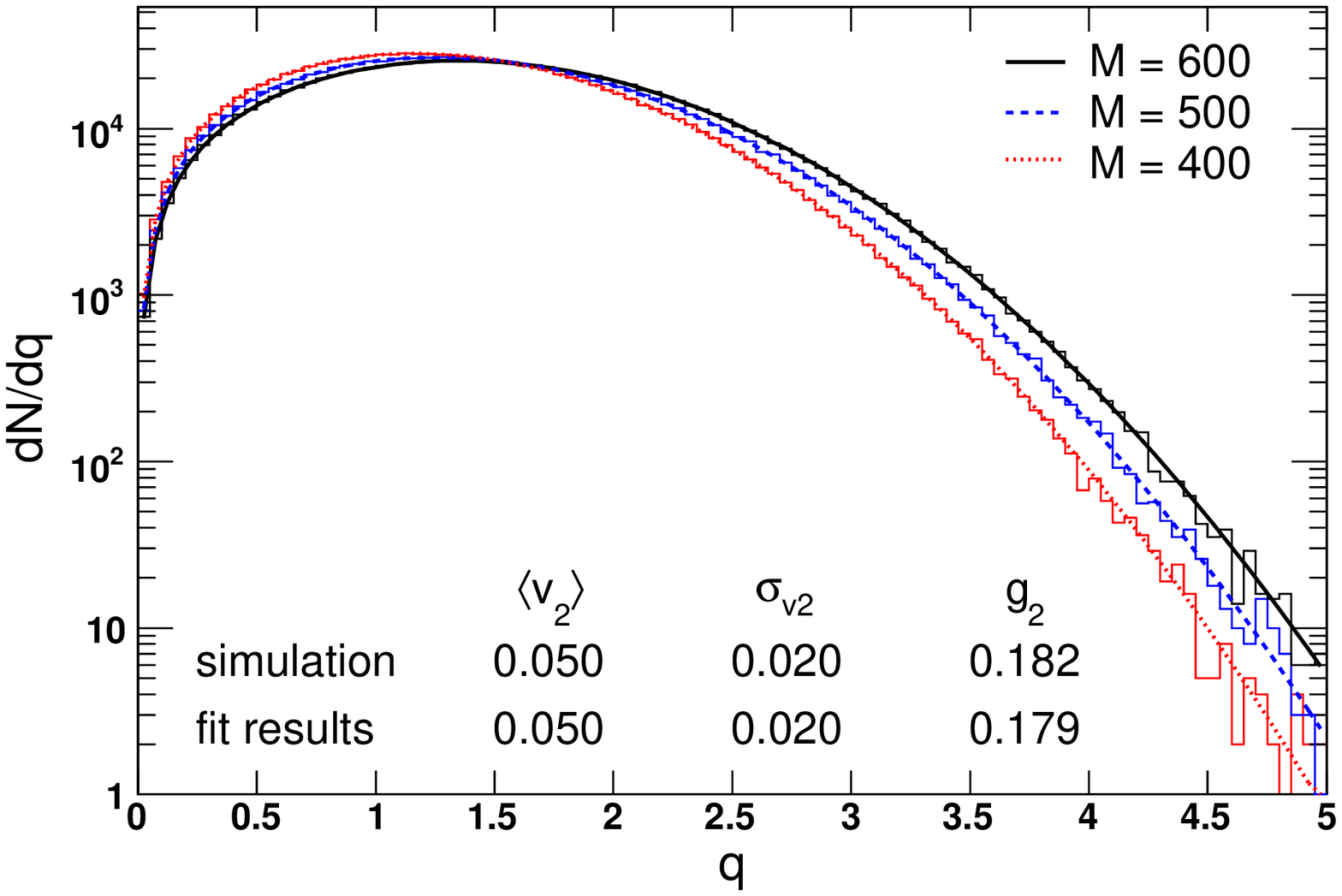}}
  \resizebox{0.425\textwidth}{!}{\includegraphics{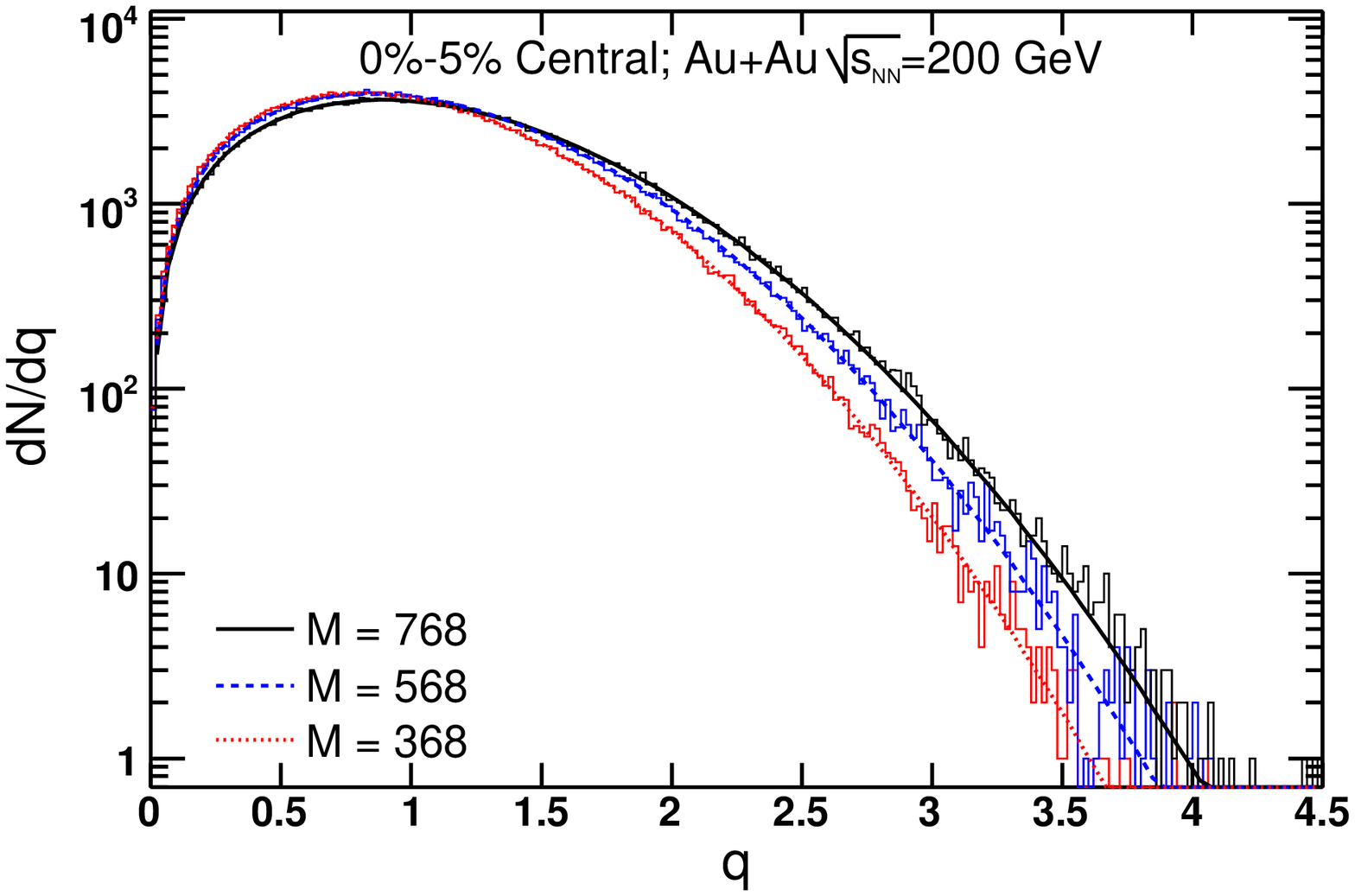}}
  \vspace{-1cm}
  \caption[]{ Simultaneous fits to three $dN/dq$ distributions
    calculated using the same event class but using different
    fractions of the total multiplicity. The left panel shows a fit to
    simulated data and the right panel shows a fit to preliminary STAR
    data. }
\label{f1}
\end{figure}

Non-flow and $v_2$ fluctuations both broaden the width of the observed
$q_2$ distribution. This ambiguity is removed by using different numbers
of tracks to calculate $q_2$. Since the $v_2$ shift of the $|q_2|$
distribution scales with $\sqrt{M}$, the broadening of the
distribution from $v_2$ fluctuations will scale with $\sqrt{M}$. We
randomly discard tracks within an event class to calculate $q_2$ with
reduced multiplicities. Then we fit the different $|q_2|$
distributions simultaneously to determine $g_2$, $\sigma_{v2}$, and
$\langle v_2 \rangle$. Typical fits are shown in Fig.~\ref{f1} for
simulated data (left) and for real data (right). Best fit values agree
well with the input parameters of the simulation.  For our data
analysis we use 15 million Au+Au collisions at
$\sqrt{s_{_{NN}}}=200$~GeV measured with the STAR
detector~\cite{STAR}.

Eccentricity (standard and participant), impact parameter $b$,
$N_{bin}$, $N_{part}$ and multiplicity are calculated from a
Monte-Carlo Glauber model~\cite{Miller:2003kd} tuned to reproduce the
multiplicity distribution of 200~GeV Au+Au collisions. We also use the
model to subtract impact parameter fluctuations. The $\sigma_{v2}$ we
present represents the width of the $v_2$ distribution at fixed
$b$. The correction is carried out according to
$\delta\sigma_{v2}=\frac{\partial
  v_2}{\partial\varepsilon_{part}}\delta\sigma_{\varepsilon,part}$, where
$\delta\sigma_{\varepsilon,part}$ is extracted from the difference between
the width of the $\varepsilon_{part}$ distribution when $b$ is allowed
to fluctuate and when it is fixed.

\begin{figure}[htb]
  \resizebox{0.50\textwidth}{!}{\includegraphics{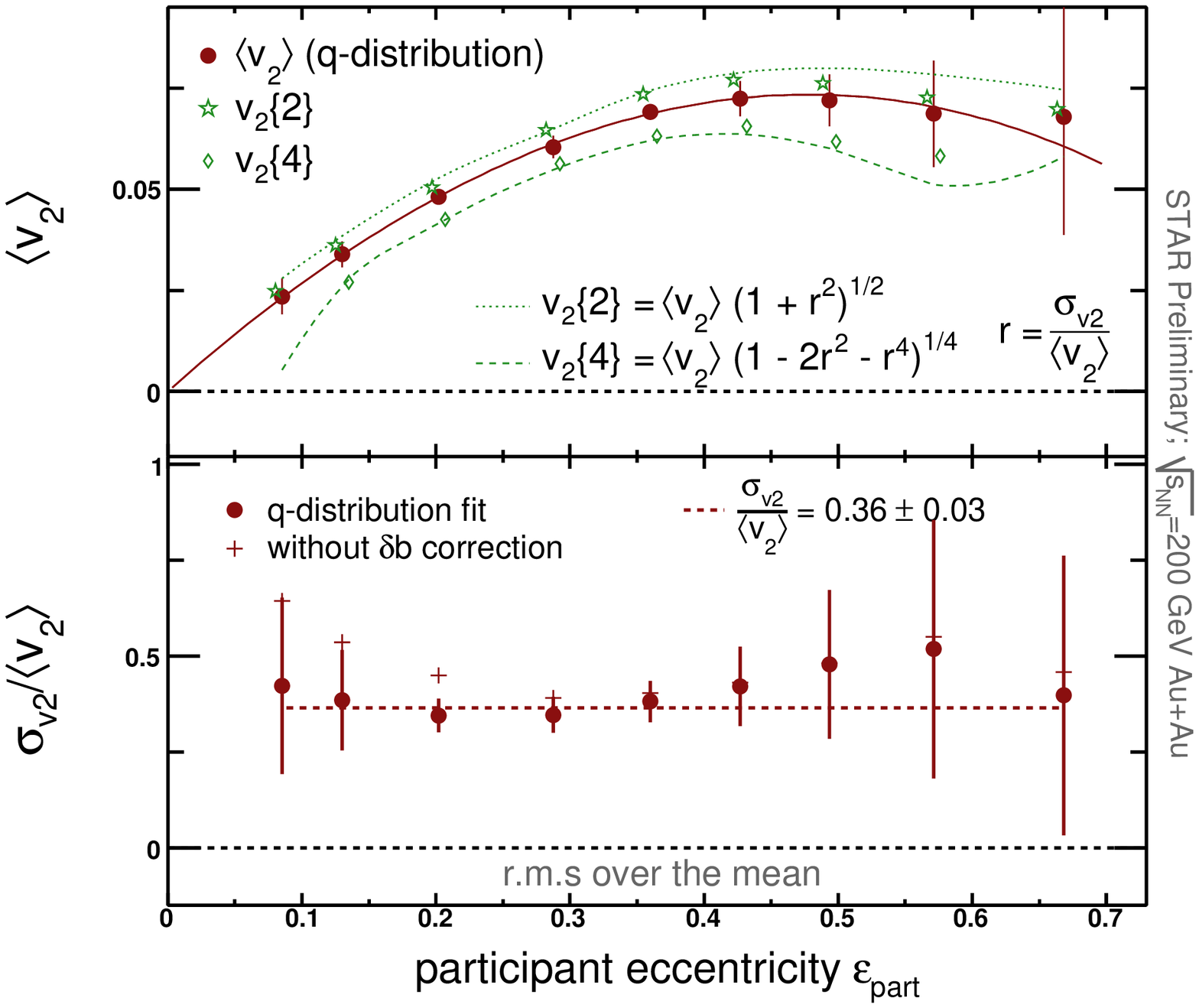}}
  \resizebox{0.50\textwidth}{!}{\includegraphics{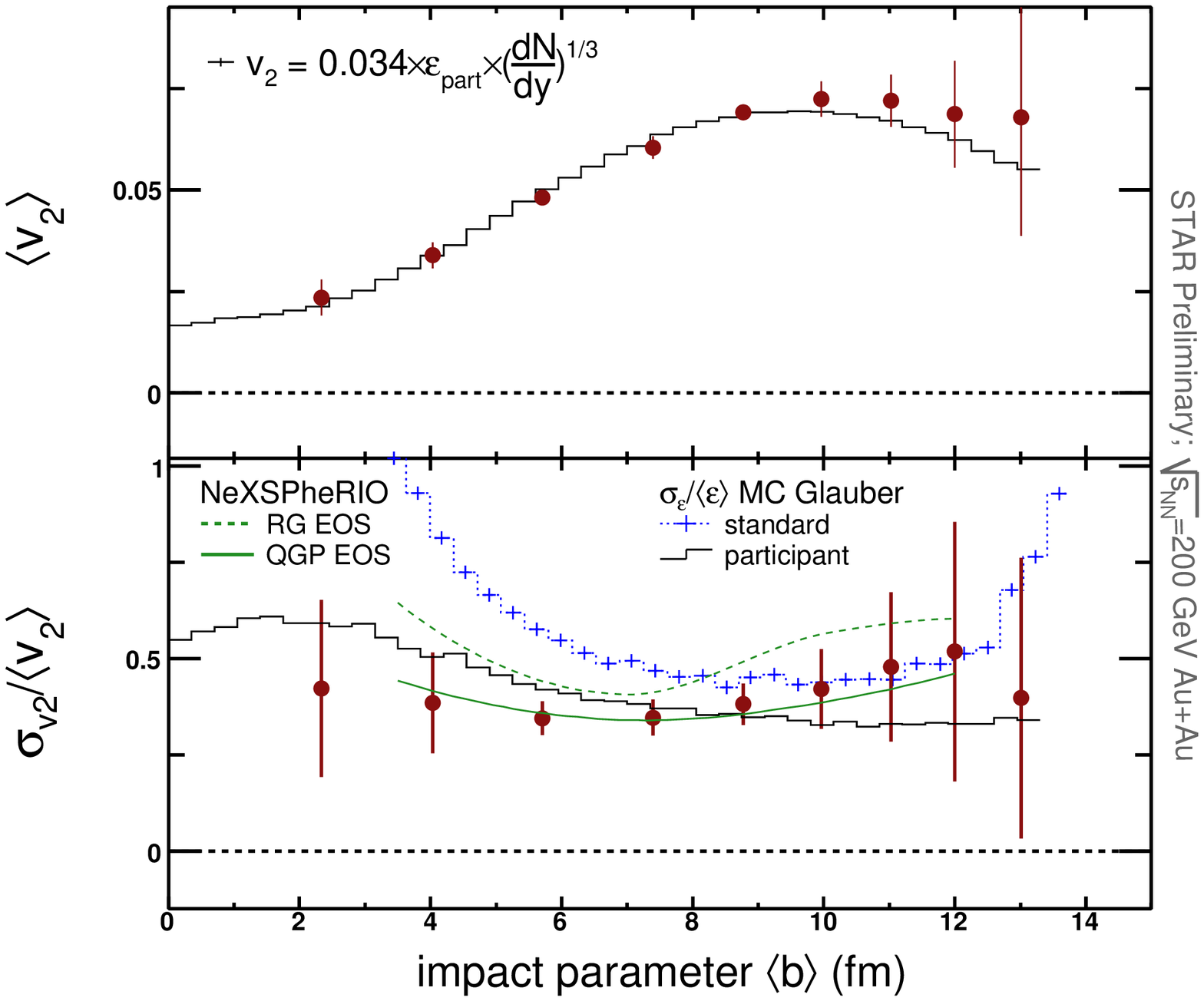}}
  \vspace{-1.5cm}
\caption[]{ The mean of the $v_2$ distribution ($\langle v_2\rangle$)
  (top panels) and the r.m.s. width of the the distribution
  ($\sigma_{v2}$) scaled by the mean (bottom panels). Data are
  presented versus participant eccentricity (left panel) and impact
  parameter (right panel). Various curves are explained in the
  text. Impact parameter fluctuations have been removed so that the
  data points represent the $v_2$ fluctuations for a fixed value of
  $b$.}
\label{f2}
\end{figure}

\textit{Results:} The $\langle v_2 \rangle$ from our fits are shown
versus $\varepsilon_{part}$ in the top panel of Fig.~\ref{f2}
(left). Measuring $v_2$ fluctuations allows us to remove the major
source of systematic uncertainty for $\langle v_2
\rangle$~\cite{Miller:2003kd,flowfl}. $\langle v_2\rangle$ projects to
zero for $\varepsilon_{part}=0$~\cite{note} giving us confidence that
our measurements are sensitive to the participant axis (a coordinate
system which can fluctuate to a different angle than the reaction
plane~\cite{epart}). We also note that $v_2 \not\propto
\varepsilon_{part}$ as would be expected from hydrodynamic
models~\cite{hydro}. The top panel of Fig.~\ref{f2} (right) shows
$\langle v_2 \rangle$ versus $b$ compared to a phenomenological model
$v_2=0.034\varepsilon_{part}\left(dN/dy\right)^{1/3}$ which is closely
related to $v_2/\varepsilon$ versus $\frac{1}{S}dN/dy$
plots~\cite{NA49}. Our data are consistent with this model.

\vspace{-0.4cm}

The r.m.s. width over the mean of $v_2$ ($\sigma_{v2}/\langle
v_2\rangle$) is shown in the bottom panels of
Fig.~\ref{f2}. $\sigma_{v2}$ is corrected for $b$ fluctuations so
that the measurements represent the width of the distribution for a
fixed impact parameter value. $\sigma_{v2}/\langle v_2\rangle$ is
approximately $0.36 \pm 0.03$ and within errors independent of
centrality.  In the top left panel, we also show a comparison of the
$\langle v_2\rangle$ and $\sigma_{v2}$ values from this analysis to
$v_2$ from 2- and 4-particle cumulant analyses ($v_2\{2\}$ and
$v_2\{4\}$)~\cite{v2papers,Miller:2003kd}. The curves derived from the
measured $\sigma_{v2}$ and $\langle v_2 \rangle$ agree with the
previously measured $v_2\{2\}$ and $v_2\{4\}$. The comparison shows
that the differences between $v_2\{2\}$ and $v_2\{4\}$ can be
explained by fluctuations without invoking non-flow suggesting that
any effects from non-flow on the integrated $v_2$ are less
important. This is not necessarily true, however, at higher $p_T$
where the ratio of $v_2\{4\}/v_2\{2\}$ decreases with $p_T$. This
decrease is likely related to an increase in non-flow correlations
from hard processes at high $p_T$.

In the bottom right panel of Fig.~\ref{f2} we compare
$\sigma_{v2}/\langle v_2\rangle$ to nucleon Monte-Carlo Glauber
calculations of $\sigma_{\varepsilon}/\langle \varepsilon \rangle$ where
the eccentricity has been calculated with respect to the reaction
plane (standard) or the participant axis (participant). The
$\varepsilon_{std}$ calculations can be excluded by data,
demonstrating that our measurements are sensitive to the initial
conditions. The relative widths of the $v_2$ and $\varepsilon_{part}$
distributions are consistent. This leaves little room for other
sources of fluctuations and suggests that the conversion of
$\varepsilon$ to $v_2$ is identically efficient for every event. This
conclusion is contradicted however by the variation of $\langle
v_2\rangle/\varepsilon$ with $\left(dN/dy\right)^{1/3}$. This
contradiction may indicate that the fluctuations are overestimated in
the Monte-Carlo Glauber model or that other sources of fluctuation are
small compared to the eccentricity fluctuations. In principle,
different treatments of the initial conditions can yield different
predictions for the $\sigma_{\varepsilon}/\langle\varepsilon\rangle$. For
example, calculations of the eccentricity based on Colored Glass
Condensate (CGC) yield eccentricity values approximately 30\% larger
than a Glauber model~\cite{cgcecc}. If the geometric fluctuations in
the CGC calculation are the same as for the Glauber model, then the
ratio $\sigma_{\varepsilon,CGC}/\langle\varepsilon_{CGC}\rangle$ will
be 30\% smaller. This would allow room for other sources of
fluctuations, consistent with the observation that $\langle
v_2\rangle/\varepsilon$ is not constant.

Hydrodynamic model calculations with fluctuating initial
conditions~\cite{spherio} are also consistent with our data when a QGP
EOS is used. The use of a resonance gas EOS seems to increase the
ratio of the rms over the mean above the observed fluctuations. This
indicates that although we have sought an observable that is sensitive
to the initial conditions, we may learn about the later evolution as
well and may even be sensitivity to the EOS. 


\textit{Conclusions:} We've disentangled non-flow and fluctuations and
extracted $\sigma_{v2}$ and $\langle v_2\rangle$ versus centrality
for Au+Au collisions at 200 GeV.  $\sigma_{v2}/\langle v_2\rangle$ is
$0.36 \pm 0.03$ and within errors independent of centrality. These
measurements significantly improve the accuracy of $\langle
v_2\rangle$, are consistent with 2- and 4-particle cumulant
measurements and demonstrate sensitivity to the initial
conditions. Future work will focus on the system-size dependence,
energy dependence, and covariance of $v_2$ for different sub-events
(\textit{e.g.} partitioned according to pseudo-rapidity and $p_T$).


\begin{thebibliography}{9}
\bibitem{note} But not necessarily for $b=0$ (Fig.~\ref{f2}, top right).
\end{thebibliography}
\end{document}